\documentclass[12pt, table]{article}

\RequirePackage{amssymb}
\RequirePackage{amsmath}
\RequirePackage{tikz}
\usepackage{xcolor}
\usepackage{url} 
\usepackage{float}
\usepackage{graphicx}
\usepackage{natbib}
\usepackage{hyperref}
\usepackage{usebib}
\newbibfield{editor}
\bibinput{ds_ethics}

\usepackage{amsthm}
\theoremstyle{definition}
\newtheorem{definition}{Definition}[section]

\newcommand{\blind}{0}

\begin{document}

\definecolor{lightgray}{gray}{0.9}
\def\spacingset#1{\renewcommand{\baselinestretch}%
{#1}\small\normalsize} \spacingset{1}


\if0\blind
{
 \title{\bf Philosophy within Data
Science Ethics Courses}
 \author{Sara Colando\hspace{.2cm}\\
  Department of Mathematics \& Statistics, Pomona College\\
  and \\
  Johanna Hardin \\
  Department of Mathematics \& Statistics, Pomona College}
 \maketitle
} \fi

\if1\blind
{
 \bigskip
 \bigskip
 \bigskip
 \begin{center}
  {\LARGE\bf Philosophy within Data Science Ethics Courses}
\end{center}
 \medskip
} \fi

\bigskip
\begin{abstract}
There is wide agreement that ethical considerations are a valuable aspect of a data science curriculum, and to that end, many data science programs offer courses in data science ethics. There are not always, however, explicit connections between data science ethics and the centuries-old work on ethics within the discipline of philosophy. Here, we present a framework for bringing together key data science practices with ethics topics. The ethics topics were collated from sixteen data science ethics courses with public-facing syllabi and reading lists. We encourage individuals who are teaching data science ethics to engage with the philosophical literature and its connection to current data science practices, which are rife with potentially morally charged decision points.
\end{abstract}

\noindent%
\vfill

\newpage

\spacingset{1.45} 
\section{Introduction}
\label{sec:intro}

Data science is a nascent field that is interdisciplinary in nature and includes an ethics component as one of the core elements. Within data science curricula, ethics courses are ubiquitous requirements for many data science majors, and there have been calls to teach more ethics in data science \citep{franklin2021,marti2023}. Recent literature has explored classes or strategies for teaching ethics in the data science context, see, for example, \citet{loux2019,shapiro2020,adhikari2021,baumer2022,noll2023}. 
As expected, however, there are diverse approaches to grappling with issues in data science ethics. 

\begin{quote}
  While there is widespread agreement that ethics must play a central role in data science education, there is less consensus on how such considerations should be delivered in the curriculum. \citep{franklin2021}
\end{quote}
One axis for incorporating ethics into a data science class or curriculum focuses on building students' understanding of the ethical issues in data science through case studies. Another axis dives deep into the philosophical conceptions and discourse surrounding key ethical terms and then connects these philosophical ideas to current data science practices. For example, data scientists work to make statistical models fair by equalizing disparate impacts between groups. 
However, not all philosophical conceptions of fairness require equalizing two groups’ outcomes. Some philosophical conceptions of fairness focus on {\em procedural equality} between individuals (e.g., see \citet{procedural-fairness}). Such discrepancies between how certain philosophers and data scientists understand equitable and ethical practices are
under-explored yet critical to achieving genuinely ethical data science practices.

While data science is a relatively new discipline, ethics is an extremely old field of study. Millennia ago, Chinese philosopher Confucius (551 - 479 BCE) and Greek philosopher Socrates (469 - 399 BCE) were independently concerned with ethical questions, like what it means to be virtuous \citep{early-ethics}. A challenge for anyone teaching data science ethics is to be accomplished in the ever-changing and growing field of data science while simultaneously being knowledgeable and well-versed in hundreds of years of ethical theory. And while, of course, no one can know everything about the two areas, we do our students a disservice if we do not acknowledge the information about, history of, and connections between the two.

In this paper, we explore the pedagogies of existing data science ethics courses using a philosophical lens. Probing the pedagogies for data science ethics is a key approach to understanding the central concepts in the field and current conversations in data science ethics research. Additionally, it also gives insights into interesting pedagogical questions such as: 

\begin{itemize}
\item Which ethics topics are most common within data science ethics classes?
\item At what stage of the data science lifecycle are the most common topics within data science ethics classes morally relevant?
\item How do we ground ethical data science questions in philosophical concepts and principles?
\item How are ethical concepts being used in data science by both data scientists and applied philosophers? 
\item How can we communicate ethical topics such that undergraduate data science students, who might not have formal philosophy training, can critically engage with them?
\end{itemize}

As a hub to compare and contrast more than a dozen data science ethics courses, we have created a \href{https://scolando.github.io/data-science-ethics/}{website}\footnote{\url{https://scolando.github.io/data-science-ethics/}}\citep{ethics_website} to address the questions above. By documenting our findings on a public website, we hope to amplify the conversation around data science ethics, the various connections between ethics topics and current data science practices, and particularly how data science ethics relates to the classroom.

In Section \ref{sec:mot}, we motivate data science ethics and provide definitions to contextualize the work done in a data science ethics course. Section \ref{sec:lifecycle} lays out the relational view of data and data models and depicts our final data science lifecycle, which we use in this paper and accompanying \href{https://scolando.github.io/data-science-ethics/}{website}. Using the syllabi and curricula from more than a dozen data science ethics courses, we lay out a novel connection between the data science lifecycle and core ethical considerations (which were common across the examined data science ethics courses) in Section \ref{sec:courses}. In Section \ref{sec:discussion}, we offer strategies for instructors who are interested in teaching data science ethics. Finally, Section \ref{sec:concl} provides encouragement for using the website we have created to inform and motivate future data science ethics courses and conversations about linking data science ethics more closely with philosophical content. 

\section{Motivation}
\label{sec:mot}

In order to motivate the pedagogical connections between data science and ethics, we start by providing frameworks for both data science and ethics. Note that there exist many different frameworks for both data science (e.g., there exist myriad Venn diagrams describing data science as the overlap between mathematics / statistics, computer science, and domain knowledge \citep{conway2011, pressman2020}) and ethics (e.g., consider {\em virtue ethics}, which emphasizes a person's virtues when evaluating the morality of an action \citep{StanVirtue} versus {\em consequentialism}, which emphasizes the consequences of the action when evaluating its morality \citep{StanConsequentialism}).  The reasons for describing the disciplines are two-fold. First, in order to communicate our ideas connecting data science and ethical topics, we need structures on which to place the topics. But more importantly, the second reason for providing the framework is to be explicit about the fact that we have made decisions about the roles that the two disciplines play in the larger data-driven decision making context. Our choices were guided by the \href{https://scolando.github.io/data-science-ethics/inside-syllabi.html}{data science ethics courses} which are at the foundation of our work (see Section \ref{sec:courses}). The work at hand seeks to characterize (not judge) the way that core philosophical tenets are being brought into data science classrooms, and our choices are motivated by the goal of describing those courses.

\subsection{What is Data Science?}
\label{data-science}

\begin{figure}[H]
\centering
\includegraphics[scale = 0.65]{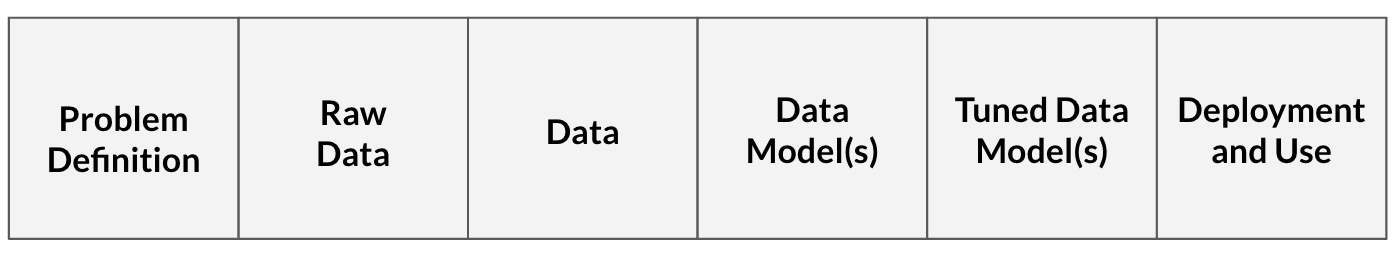}
\caption{Processes involved in data science. The amount of processing required to get the outputted object (e.g., raw data, data, a tuned model, etc.) increases as we move from left to right.}
\label{fig:ds_process}
\end{figure}

\begin{definition}[Data Science]
Data science courses are often focused on transforming data into data model(s), but data science as a field encompasses all the processes needed to answer questions with data: from ``Problem Definition" to ``Deployment and Use" in Figure \ref{fig:ds_process}. 
\end{definition}

Below, we walk through the data science processes described in Figure \ref{fig:ds_process} with a hypothetical example to illustrate what happens at each processing step and how these steps relate to one another.

\begin{itemize}
\item {\bf Problem Definition:} defines the question we aim to answer with data. We answer questions like what counts as ``success" (i.e., when do we say a data model is successful), and how can we actually measure (or approximate) our event of interest?

\begin{quote} E.g., suppose we aim to answer: what is the likelihood that a customer purchases product S? Only when our model (built on training data) achieves an overall accuracy $\ge$ 0.75 (for the test data) do we say that our model is successful.
\end{quote}

\item {\bf Raw Data:}
the information collected from interactions with the world.

\begin{quote} E.g., our raw data includes information from each time the customer clicks on an advertisement for product S, including the timestamps for each advertisement interaction and the customer's demographic information.
\end{quote}

\item {\bf Data:} 
the processed form of the raw data. This often is the tabularized version of the raw data.

\begin{quote} E.g., a data table where each row represents a unique customer and the columns are the variables that describe that customer. These columns include information from the raw data (e.g., the number of times they click on an advertisement for product S, their age, etc.) and also any engineered variables (e.g., their average time between advertisement clicks). During data generation, we also decide what to do with any missing values (i.e., do we leave them, ignore them, or impute them?).
\end{quote}

\item {\bf Data Model(s):}
the product created from running the input data through a learning algorithm (i.e., a mathematical formula that predicts an output for a given input). Data Models aim to generalize the relationship between variables in the data. 

\begin{quote} E.g., we choose to use logistic regression, where our response variable is the binary indicator of whether or not the customer purchased product S. Our explanatory variables are the customer's age, their number of advertisement clicks, and the average time between advertisement clicks. The general form of our data model would be:
\end{quote}
\begin{eqnarray*}
\textnormal{logit}(p(\textnormal{purchase S})) = \beta_0 & + &\beta_1 \cdot \textnormal{age} \ + \ \beta_2 \cdot \textnormal{ad-click number} \ + \nonumber \\
&&\beta_3 \cdot \textnormal{average time between ad-clicks}
\end{eqnarray*}

\item {\bf Tuned Data Model(s):}
a data model in which the model's parameters (including hyperparameters) are adjusted, usually to better balance the model's generalizability and (prediction) accuracy for the population of interest. In practice, tuning is typically done by either splitting the data into training and test sets or by performing cross-validation.

\begin{quote} E.g., we choose to do 5-fold cross-validation to tune our data model's parameters in order to maximize our model's overall accuracy at a particular cutoff value. With logistic regression, the model tuning comes as a choice of which variables to include (because logistic regression does not have any hyperparameters).
\end{quote}

\item {\bf Deployment and Use:}
generation of predictions (or other output) from the tuned data model. This is the stage where we ask questions like where should the system be used, who should be using it, and who/what should the data model be used on?

\begin{quote} E.g., we ultimately decide that we should only use our tuned data model on US customers who are under a certain age. Also, we determine that only data scientists at the company who are selling product S should be able to access the data model and generate predictions with it.
\end{quote}
\end{itemize}

\subsection{What is Ethics?}
\label{sec:intro-ethics}

\begin{definition}[Ethics] Ethics is comprised of three main branches:

\begin{enumerate}
  \item {\bf Applied Ethics} concerns the treatment of ``moral problems, practices, and policies in personal life, professions, technology, and government," \citep{applied-ethics}. 

\begin{quote} E.g., Should we ever deploy predictive policing algorithms? If there is a shortage of ventilators, who should get one?
\end{quote}

\item {\bf Ethical Theory} concerns ``the articulation and justification of the fundamental moral principles that govern how we should live and what we ought to morally do," \citep{ethical-theory}. Types of overarching ethical theories include {\em consequentialism} \citep{StanConsequentialism}, {\em deontological ethics} \citep{StanDeontological}, and {\em virtue ethics} \citep{StanVirtue}. It is important to note that many ethical theories are very abstract. As such, ongoing philosophical work is committed to offering, critiquing, and translating abstract ethical theories into advice about how we should actually live and what is morally permissible in practice. It is also worth noting that some philosophers just focus on what we should do in specific cases and do not appeal to overarching ethical theories at all.
 
\begin{quote}E.g., Why should I be just? What constitutes respecting others? 
\end{quote}

\item {\bf Metaethics} explores ``the status, foundations, and scope of moral values, properties, and words," \citep{metaethics}.

\begin{quote} E.g., When we say an action is morally ``wrong", do we mean the action has a certain feature that is bad, that I have a negative feeling towards the action, or something else? How do we come to know whether moral claims (e.g., we should respect others) are true or false?
\end{quote}

\end{enumerate}
\end{definition}

A key takeaway is that ethics is concerned with {\em normative} questions. For instance, ethics attempts to answer not what person A morally values, but rather what person A {\em should} morally value. Similarly, ethics is interested in what decision person A {\em should} make in a given context and not what decision person A actually makes (or is likely to make) in that context. 

Another important distinction is between ethics and the law. Actions can be legal without being morally permissible. Conversely, actions can be illegal but still morally permissible. For example, it seems morally impermissible to plagiarize a paper in school even though it is not against the law to do so. As such, we cannot simply appeal to the law in order to understand what is morally (im)permissible. Rather, we need to defend and appeal to moral principles. For instance,  we might defend that plagiarizing a paper is morally impermissible by arguing that it is deceptive. That is, plagiarism misrepresents another person's work as one's own without properly crediting them (e.g., citing them). However, is deception always morally impermissible? Suppose a person deceives their friend about their whereabouts to keep the friend's surprise birthday party a secret. In this case, the friend's deception does not seem morally impermissible. What, then, if not merely being a case of deception, explains the intuition that plagiarizing a school paper is morally impermissible? Ethics aims to answer such questions and, more broadly, offer a methodical way of approaching such questions and, with that, making morally good decisions.

\subsection{What is Data Science Ethics?}
\label{subsec:data-science-ethics}

Data science ethics is usually considered a subfield of applied ethics. However, case studies in data science ethics can also be used to explore questions in ethical theory and metaethics.

\begin{definition}[Data Science Ethics]

Per \citet{DS-ethics-def}, data science ethics studies and evaluates moral problems related to:
\begin{itemize}
  \item {\bf Data} including generation, recording, processing, dissemination, and sharing
  \item {\bf Algorithms} including artificial intelligence, machine learning, large language models, and statistical learning models
  \item {\bf Corresponding practices} including responsible innovation, programming, hacking, and professional codes\footnote{In the context of data science ethics, professional codes are guidelines that outline the ethical standards and responsibilities for those engaged in data science work. Many companies and organizations have produced professional codes related to data science ethics, such as \href{https://www.microsoft.com/en-us/ai/responsible-ai?ef_id=_k_CjwKCAjw65-zBhBkEiwAjrqRMOOu4pQdMiA-H3F4IAPwAFy6P5AteC8WR1R1ry6SETKE3Zhlhgi4ABoC5HIQAvD_BwE_k_&OCID=AIDcmm1o1fzy5i_SEM__k_CjwKCAjw65-zBhBkEiwAjrqRMOOu4pQdMiA-H3F4IAPwAFy6P5AteC8WR1R1ry6SETKE3Zhlhgi4ABoC5HIQAvD_BwE_k_&gad_source=1&gclid=CjwKCAjw65-zBhBkEiwAjrqRMOOu4pQdMiA-H3F4IAPwAFy6P5AteC8WR1R1ry6SETKE3Zhlhgi4ABoC5HIQAvD_BwE}{Microsoft}, \href{https://www.ibm.com/watson/assets/duo/pdf/everydayethics.pdf}{IBM}, and the \href{https://www.unesco.org/en/artificial-intelligence/recommendation-ethics}{United Nations}.}
\end{itemize}

\end{definition}

A famous case study that highlights the moral problems related to data science is the Correctional Offender Management Profiling for Alternative Sanctions algorithm (COMPAS). COMPAS generates a risk score for each defendant based on their predicted likelihood of being convicted. This risk score is then used to inform decisions in the United States criminal justice system (e.g., to set bond amounts, determine criminal sentencing, and decide early release for parole). In 2016, ProPublica researchers found that Black defendants were twice as likely as white defendants to be falsely labeled as recidivists by COMPAS \citep{COMPAS}. Additionally, white defendants were more likely to be mislabeled as having a lower risk of recidivism than Black defendants \citep{COMPAS}. As such, \citet{COMPAS} concluded that COMPAS was unfair to Black defendants. However, other literature complicates what it means for an algorithm to be {\em fair}. For instance, some argue that fairness requires {\em predictive parity}, which in the case of COMPAS means that if Black and white defendants were each given the same risk score, they would be equally likely to recidivate \citep{2016COMPAS}.

Yet, researchers have found that when base rates are different, as they are for recidivism across Black and white defendants in the United States, an algorithm cannot simultaneously satisfy {\em equal false positive rates}, {\em equal false negative rates}, and {\em predictive parity} across groups \citep{kleinberg2016, chouldechova2017, Corbett-Davies}. This has prompted discussions about what is actually required for algorithmic fairness and whether statistical criteria, like {\em equal false positive rates}, {\em equal false negative rates}, and {\em predictive parity}, actually track important aspects of the normative concept of fairness.\footnote{See \citet{fairness-book} for an overview of fairness measures in machine learning.} Thus, we need to understand the normative concept of fairness in order to assess whether COMPAS is unfair to Black defendants and, moreover, whether it is even possible in principle for COMPAS to be fair given the unequal recidivism base rates across Black and white defendants in the United States.

\subsection{Importance of Data Science Ethics}
\label{imp-ds-ethics}

\begin{figure}[H]
\centering
\includegraphics[scale = 0.55]{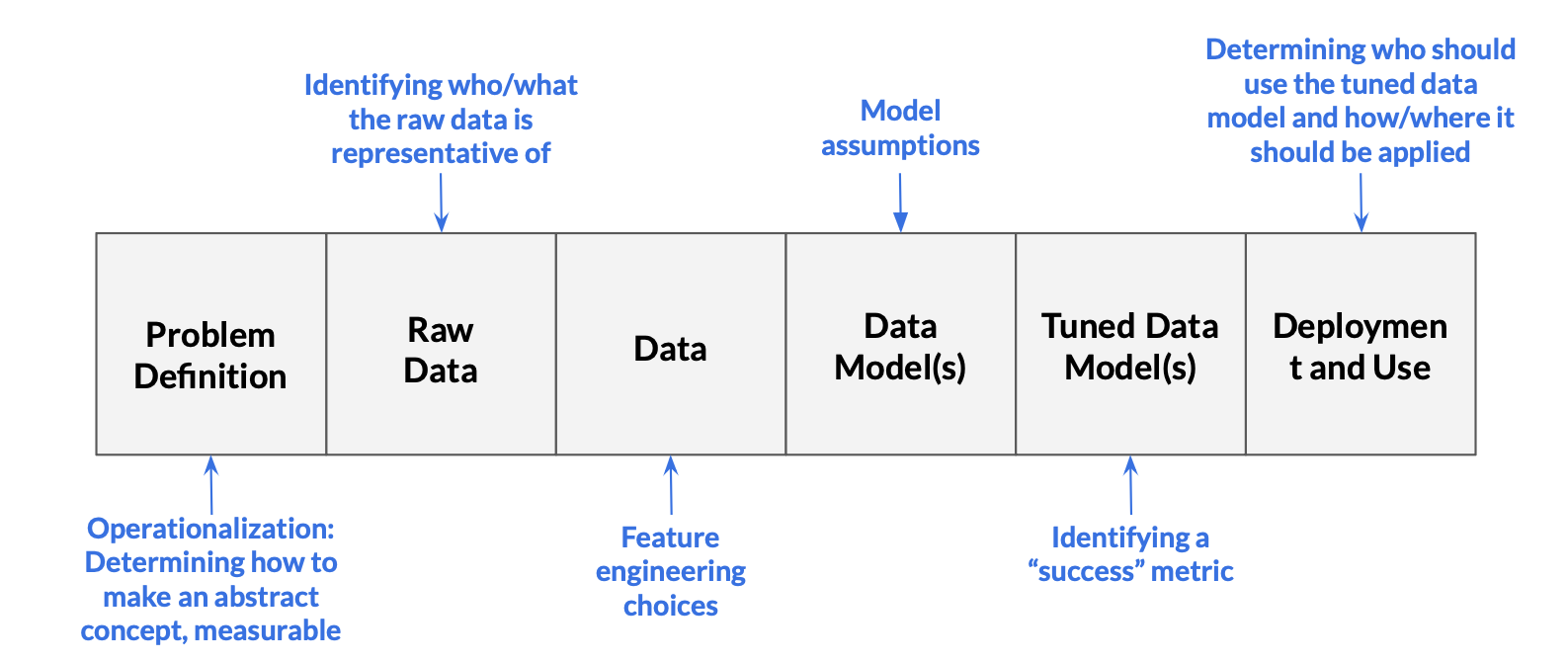}
\caption{Some key decision points that could produce morally charged outcomes have been added to the data science processes diagram. A data scientist may or may not be aware of these decision points or responsible for all these decisions in practice.}
\label{fig:DS-stack-values}
\end{figure}

There are a number of critical decision points in data science, which can lead to moral problems in data, algorithms, and corresponding practices. Figure \ref{fig:DS-stack-values} connects potentially morally charged decision points with data science processes.

Given the commonly held belief that mathematics, and consequently statistics, is objective in the sense that it is not influenced by factors such as the practitioner's moral values, potentially morally charged choices in data science are often made implicitly, without the decision-maker reflecting on, for example, how opting one choice over another (mis)aligns with their best judgment about what they ought to do or their moral duties to stakeholders. The ethical considerations at data science decision points must be made explicit: both the existence of a choice and the moral implications of the practitioner's ultimate decision are key aspects of each data science stage.

Data science ethics aims to illuminate the moral implications of choices within data science and takes an interdisciplinary perspective on aligning our data science practices with what we ought to do and our moral duties to stakeholders. For example, returning to the COMPAS example, data science ethics would address questions such as: what does it mean for an algorithm to be fair or just? Does algorithmic fairness or justice require satisfying some statistical criteria, and if so, which one(s)? Does Equivant, the company that made COMPAS, have a duty to make their algorithm transparent, explainable, or even fair? Engaging with such questions, and with data science ethics more generally, is critical to ensuring morally permissible data science practices. This engagement is particularly important given we live in the age of Big Data, where decisions with high moral stakes, like pretrial release \citep{COMPAS}, home loan approvals \citep{homeloan-biased}, and Child Protective Service's welfare visits \citep{ChildWelfare-biased}, are increasingly being influenced by data science.

\section{Data Science Lifecycle}
\label{sec:lifecycle}

To take an interdisciplinary perspective on aligning what we should do and our moral duties to stakeholders with our actual data science practices, we must understand where common ethics topics emerge in data science. This requires us to choose a data science lifecycle that characterizes how interactions with our world are transformed into data models. Though it might not be explicit, choosing one lifecycle over another endorses specific views about data, data models, and their respective relationships to what we interpret as knowledge about our world. Thus, the choice to use a certain data science lifecycle is value-laden, an important point to reiterate to students.

Although slightly beyond the scope of the current paper, we describe the representational view of data and data models in Appendix \ref{app-rep}. We include the representational view for those instructors who believe that contrast is one way to understand an idea. Below, we describe the relational view of data and data models, a framework that is likely very familiar to statisticians and data scientists.

\subsection{The Relational View of Data and Data Models}
\label{relational}

Under the relational view of data and data models, data is understood as an object that is treated as evidence for a claim about the world and that can be circulated amongst individuals or groups \citep{Leonelli}. The informational content of data depends on the researchers’ background assumptions and social context. So, information is not inherent to the data but is instead defined by the social environment in which the data is collected and the function the data is supposed to serve. The view that context plays an integral role in what data represents is also emphasized within the statistics community. As statisticians George Cobb and Thomas Moore famously claimed, ``[data] are numbers with a context" \citep{Cobb-Moore}. Given data is context-sensitive in the sense that its informational content is influenced by researchers' assumptions and social contexts, data models (rather than the data itself) are taken to represent relationships in our world under the relational view. Thus, the relational view of data and data models highlights that data models are a necessary and highly influential aspect of what we take to be knowledge about our world in data science \citep{Leonelli}.


\subsection{The Final Data Science Lifecycle}
\label{subsec:final_lifecycle}

\begin{figure}[H]
\centering
\includegraphics[scale = 0.7]{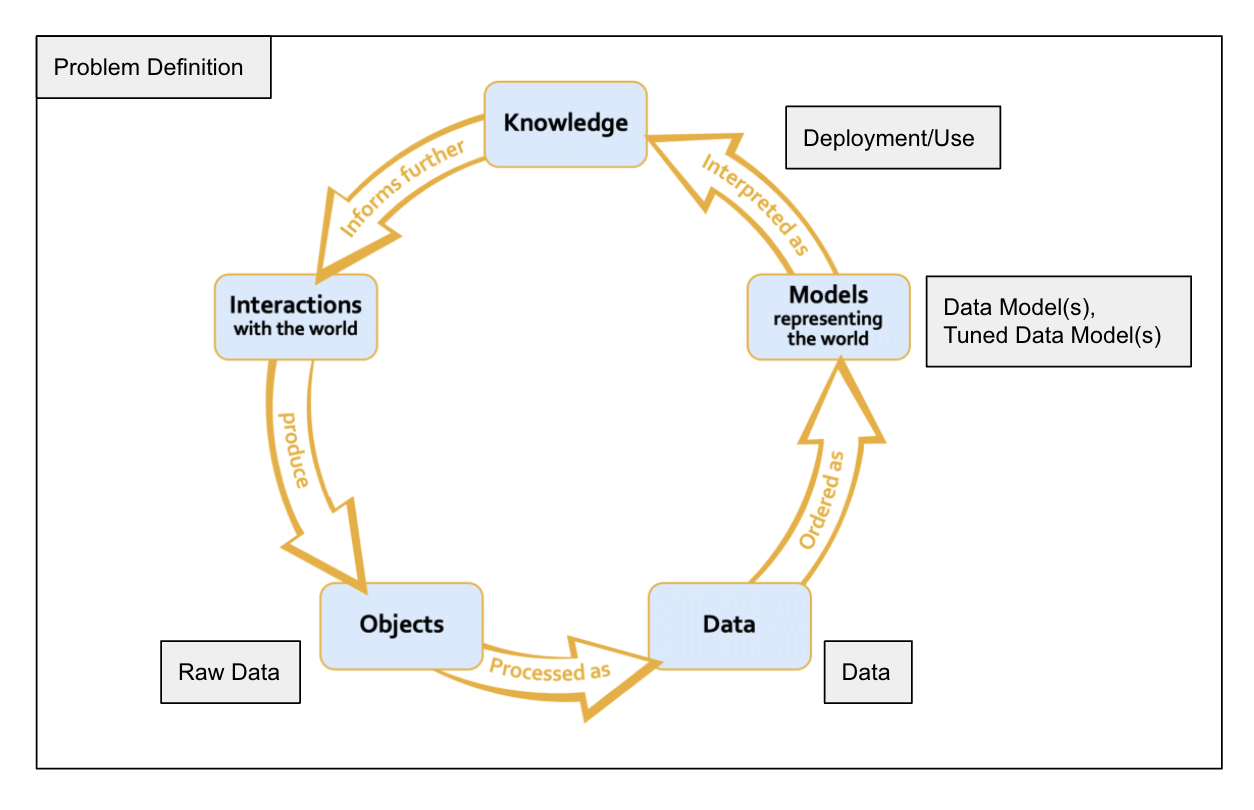}
\caption{Data science lifecycle under the relational view of data and data models, superimposed with the data science processes, described in Section \ref{data-science}. ``Raw Data" is equivalent to objects, which is processed into ``Data". Data Model(s) result from ordering data in a particular way (e.g., as a linear regression). ``Tuned Data Model(s)" also result from ordering data. However, tuned models usually involve {\em more} data (e.g., additional data collected because the data model(s) was unsuccessful with respect to some chosen metric or {\em different} data (e.g., test data) than the original data model(s). ``Deployment/Use" involves interpreting knowledge from the model(s). ``Problem Definition" affects the entire lifecycle, which is why the ``Problem Definition" box surrounding the entire lifecycle (e.g., how we define the question we want to answer with data and model success influences what raw data we process and how we tune our model). The layout for this figure was inspired by the diagram on pg. 58 of \citet{Beaulieu-Leonelli}.}
\label{fig:lifecycle-practices}
\end{figure}

Given that the relational view aligns well with data science examples (see Appendix \ref{rep-rel-in-practice}), we use the data science lifecycle in Figure \ref{fig:lifecycle-practices} throughout this paper and on the associated website. Beyond aligning well with data science examples, the lifecycle shown in Figure \ref{fig:lifecycle-practices} also fits with the data science processes described in Section \ref{data-science}. Figure \ref{fig:lifecycle-practices} showcases how the data science processes connect to our final data science lifecycle. However, other data science lifecycles could also be compatible with the relational view of data and data models as well as the described data science processes (we outline some potential shortcomings of our final lifecycle in Section \ref{flaws-lifecycle} for instructors who want to have their students walk through the pros and cons of working with different versions of the data science lifecycle). As such, we acknowledge that our selected lifecycle, which is depicted in Figure \ref{fig:lifecycle-practices}, is not the only viable choice.

\section{Data Science Ethics Courses}
\label{sec:courses}

The impetus for our work was a survey of sixteen data science ethics courses. We started by doing a literature review using the syllabi and reading lists for the courses, and we soon realized that there are many interesting connections to investigate across the courses. In what follows, we connect the syllabus topics related to ethics with the data science lifecycle. 

\subsection{The Courses}

The sixteen courses we investigated are \href{https://scolando.github.io/data-science-ethics/inside-syllabi.html}{described in detail} on our \href{https://scolando.github.io/data-science-ethics/}{website}. They were chosen by exploring the curricula of notable data science programs. We focused only on those programs that contained public-facing syllabi and reading lists. There were eight courses that were philosophy-oriented (i.e., the course's reading lists and student assignments were more philosophical in nature, e.g., the authors on the course's reading list were mostly philosophers, and students had to write philosophical papers), four that were data science-oriented (i.e., the course's reading lists and student assignments were more technical in nature, e.g., involved preprocessing approaches to debiasing datasets), and four that could not be easily categorized into either group. Most courses did not have prerequisites, and they were a mix of undergraduate and graduate courses. The faculty teaching the courses came from a large range of disciplines, including philosophy, data science, statistics, computer science, engineering, politics, history, and law. Six of the courses were joint-taught by professors from different disciplines. The courses are taught at small and large schools, both public and private. After parsing through the syllabi and assigned readings, we came up with the following groups of topics related to ethical data science, see Table \ref{tab:topics}.

\begin{table}[H]
\begin{center}
\rowcolors{1}{}{lightgray}
  \begin{tabular}{|l|c|}
\hline
Syllabus Topic & \# of courses\\
\hline
Privacy & 11\\
Bias & 8\\
Fairness & 8\\
Explainability & 6\\
Workplace & 5\\
Alignment & 4\\
Transparency & 4\\
Causation & 3\\
Characterizations of Data and Data Science & 3\\
Consent & 3\\
Democracy & 3\\
Interpretability & 3\\
Justice & 3\\
Predictive Policing & 3\\
Responsibility & 3\\
\hline
\end{tabular}
\caption{The most frequent ethics topics across the 16 data science ethics courses considered in our study.}
\label{tab:topics}
\end{center}
\end{table}

We determined each syllabus' topics by extracting 1-5 keywords from the course topic list, which were mentioned either explicitly or implicitly on the syllabus (on the listed course outcomes, reading synopses, etc.). We then parsed through these keywords and grouped them under broader topics. For example, ``Labor Rights" and ``Workplace Surveillance" were both grouped under the keyword of ``Workplace". The most frequent topics were the keywords that appeared at least three times.

In what follows, we explicitly connect the most frequent syllabus ethics topics and data science topics via the data science lifecycle provided in Section \ref{subsec:final_lifecycle} to demonstrate the moral dimensions of various steps in the data science lifecycle. The connections are further explored via quintessential or typical examples (e.g. how bias can arise in interactions with the world) as well as through less conventional examples (e.g., how bias can also arise when imputing missing values during data processing). The quintessential or typical examples were determined based on which connections were provided in articles from the collated syllabi's reading lists.

\subsection{Lifecycle Locations of the Most Common Syllabi Topics}

\begin{figure}[H]
\centering
\includegraphics[scale = 1.25]{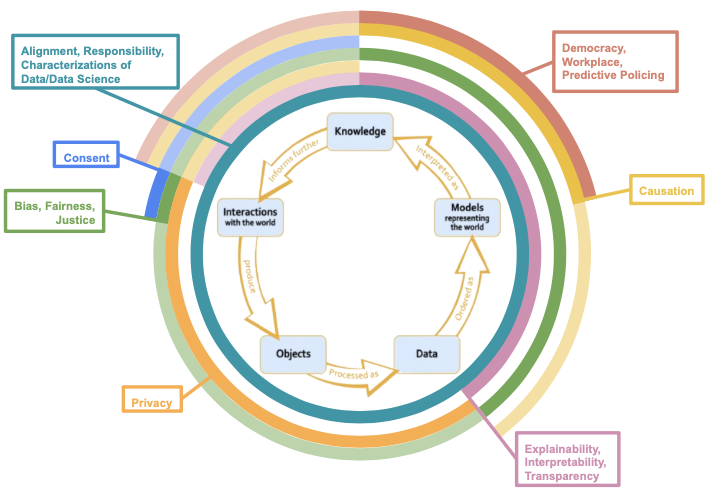}
\caption{Data science lifecycle overlayed with concentric circles. The opaque concentric circles represent paradigmatic connections between the common ethics topic from the course syllabi and the data science stages. The transparent arcs denote areas where there is a connection between the common syllabi topics and the data science stage, but the connection is less conventional. }
\label{fig:lifecycle_2}
\end{figure}

In Figure \ref{fig:lifecycle_2}, opaque rings represent paradigmatic\footnote{A paradigmatic example is a typical or quintessential example of a topic or concept. E.g., simple linear regression would be considered a paradigmatic example of a data model.} ethical issues that relate to each of the most frequently mentioned topics in the data science ethics syllabi. They are aligned with the data science lifecycle to demonstrate the important connections between ethics and data science. Transparent rings denote areas where there is still substantial overlap between frequent ethics topics and the data science stage, but the example is not a conventional application of the ethics concept to data science.


\subsection{Connections}
\label{sec:connections}

In this section, we offer paradigmatic and less conventional examples that demonstrate how the most frequent ethics topics from the syllabi arise throughout the data science lifecycle. The subsections are ordered from the first paradigmatic connection between the ethics topic and a data science stage, starting from ``interactions with the world" and then proceeding counterclockwise around the data science lifecycle. When understanding why ethical issues arise and are salient in a particular case, it is often beneficial to reflect on our intuitions in the case (e.g., that plagiarizing a school paper is morally impermissible). However, as also seen in the plagiarism case, explaining our intuitions involves developing a well-defended moral principle which often requires us to critically examine the validity of various moral principles (e.g., that deception is always morally impermissible) and then either scrap or revise the moral principles accordingly. Given the nuances of the ethical issues that arise in data science and ongoing discussions in philosophy about them, we direct the reader to our website’s reading lists (linked in each subsection's header) to further explore the intricacies and applications of the ethics topics to data science.

\subsubsection{\href{https://scolando.github.io/data-science-ethics/readings/Characterizations.html}{Characterizations of Data and Data Science}}

As mentioned in Section \ref{sec:lifecycle}, the choice of how to characterize data and data science is value-laden as it reflects a particular interpretation of knowledge from data and data models. For instance, if we view data as a direct representation of the world, we might overlook biases or other ethical issues that arise during the collection or data processing stages. On the other hand, if we view data and data models as context-dependent, then we can acknowledge that ethical issues, like algorithmic bias, can arise during the data collection and processing stages. In turn, characterizations of data and data science influence every stage of the data science lifecycle.

\subsubsection{\href{https://scolando.github.io/data-science-ethics/readings/Alignment.html}{Alignment}}

Alignment focuses on whether (and if so, how) our moral values are reflected in our current data science practices. Given that potentially morally charged decision points exist throughout data science (see Section \ref{imp-ds-ethics}), alignment is also an important ethics topic at every stage of the data science lifecycle.  

\subsubsection{\href{https://scolando.github.io/data-science-ethics/readings/Responsibility.html}{Responsibility}}

While both moral and legal responsibility are important considerations in data science, we focus here on moral responsibility. Again, an individual might be morally responsible for their behavior, even if they would not be legally responsible for it (e.g., an individual might be morally responsible for plagiarizing a school paper but not legally responsible for it since the act of plagiarism is not against the law). 

\paragraph{Paradigmatic example(s):} Discussions of moral responsibility in data science typically concern model deployment and ``interactions with the world". For example, Amazon was held morally responsible for deploying a hiring algorithm that was biased against female applicants \citep{Amazon}. Moral responsibility also arises when building data models. For instance, it seems reasonable to contend that if another company had created Amazon’s faulty hiring algorithm, then that company would also be morally responsible, alongside Amazon, for the biased results if they did not adequately define where the model should be used or who should use it. 

There are also several case studies (in data science and beyond) where people are morally responsible for failing to obtain informed consent when collecting personal information (i.e., during ``interactions with the world"). Some examples include the collection of HeLa cells\footnote{In 1951, cervical cancer cells from Henrietta Lacks were taken without her knowledge. Her cells and cell line, known colloquially as HeLa cells, have been widely used in science ever since \citep{HeLa}.} and the commercialization of social media users' data without obtaining their informed consent.\footnote{See \citet{social-media-example} for issues with informed consent and social media.}

\paragraph{Less conventional example(s):} Though less commonly thought about, moral responsibility is also relevant during data processing (i.e., between ``interactions with the world" and ``data"). Specifically, it seems reasonable to hold data scientists morally responsible for the moral harms that arise from their data cleaning or storage practices. For example, if a data scientist stored personal data in a foreseeably faulty database, they would be at least partially morally responsible for any data leakages. Similarly, if the data scientist who was supposed to remove identifiers from the data was negligent, they would have (at least some) moral responsibility for the ethical repercussions that arise from the data not being adequately anonymized.\footnote{See \citet{OkCupid} for an example of how moral responsibility can be relevant when failing to remove identifiers from data.} Or yet when performing other processing, like deciding to code age as continuous or ordinal, the data scientist is responsible for their analysis choices.

\subsubsection{\href{https://scolando.github.io/data-science-ethics/readings/Consent.html}{Consent}}

\paragraph{Paradigmatic example(s):} Generally, when people think about informed consent in data science, they think about it during data collection (i.e., during ``interactions with the world"). That is, they consider whether the researcher or company has gathered informed consent when collecting people's data. When researchers or companies get permission to collect people's data, they also typically ask for consent to use it in a specific capacity later in the data science lifecycle (e.g., to build data models, store it in a database, or share with another company) given that obtaining a person's informed consent is often essential for respecting their autonomy \citep{sep-informed-consent}.

\paragraph{Less conventional example(s):} While less conventional, issues of consent also arise when applying insights from data models to future interactions with an individual, even if none of the individual's data was used to build (or test) the data model. For example, suppose that a job site creates a data model that predicts that people from a certain demographic group are more likely to interact with a nannying job post than a construction job post. A new person from that demographic group then engages with the job site. Based on the predictive model, the social media platform shows the nannying job post to them instead of the construction job post.\footnote{This example is partially inspired by the examples in \citet{hiring-example}.} It seems important for the social media company to obtain the new user's informed consent to use the predictive model on them, given that the predictive model undermines their autonomy in some capacity. Namely, using the data model on the new user restricts them from seeing that certain jobs are available.\footnote{See \citet{sep-informed-consent} for reasons beyond respecting a person's autonomy why obtaining informed consent is morally important.}

\subsubsection{\href{https://scolando.github.io/data-science-ethics/readings/Privacy.html}{Privacy}}

\paragraph{Paradigmatic example(s):} Often, privacy is connected to data science between interactions with the world and data processing. For example,  university researchers posted profile data from the OkCupid dating site to an open data repository in 2016 \citep{OkCupid}. The data revealed intimate details about more than 70,000 users, including their usernames, sexual preferences, and personal opinions \citep{OkCupid}. Around 30\% of the profiles were identifiable, meaning that those profiles could be connected to their real name \citep{OkCupid}. The researchers' violation of the OkCupid users' privacy seems morally problematic. As explained in \citet{sep-privacy}, one reason why violating users' privacy is morally problematic in this case is that it endangers the users' abilities to control their relationships with others. That is, privacy is a way of ``modulating" our degrees of friendship with others, i.e., we would share more personal details about our life with someone we are better friends with \citep{sep-privacy}. Additionally, some philosophers contend that the right to privacy is grounded in the right to autonomy (e.g., rights over one's personal property and own body) \citep{sep-privacy}.  




\paragraph{Less conventional example(s):} Worries about privacy can resurface when using insights from a data model to inform our future interactions with the world. For instance, suppose that a company's data model predicts that an applicant is unqualified for a job. The company decides to share this prediction with a list of other major employers. There is an intuitive sense in which publicizing the model's prediction to several other employers seems morally problematic, given that it entails sharing personal information about the applicant without their consent. However, would it be unethical for a boss at one company to share their belief that the person is unqualified for a job with their friend, who is a boss at another major company? It seems to be less morally tenuous even though the friend sharing their evaluation of the applicant with their friend still entails sharing personal information about the applicant without their consent. To explain our difference in intuitions in these two cases, we need a well-defended moral principle about what grounds someone's right to privacy.

One justification for our difference in intuitions between the two cases is that there is a difference in scalability and damage in the algorithm versus the friend case. In her book, {\em Weapons of Math Destruction}, Cathy O'Neil breaks down different algorithms in terms of their opacity, scalability (pernicious feedback loops), and damage (ability to grow exponentially). She puts forward a moral principle to determine whether an algorithm is a {\em Weapon of Math Destruction}, in which case, she argues, it should not be used \citep{oneil2016}. Avoiding pernicious feedback loops is one such strategy to ground privacy concerns and explain the difference in intuitions between the algorithm (which has a scalable pernicious feedback loop) and the friend case (which doesn't). 

\subsubsection{\href{https://scolando.github.io/data-science-ethics/readings/Bias.html}{Bias, Fairness, Justice}}

\paragraph{Paradigmatic example(s):} There are many cases of biased, unfair, and unjust data models. One such example is Amazon's now-scrapped hiring algorithm, which was critiqued for being biased against female applicants \citep{Amazon}. The algorithm’s goal was to identify ‘ideal candidates’ for technical positions at Amazon from hundreds of resumes submitted. Here, `ideal candidates’ were considered those whose resumes were most similar to resumes that had been previously submitted to Amazon over the last ten years. However, given that the technology industry is heavily male-dominated, most of these resumes came from male applicants, and consequently, the algorithm’s predictions turned out to be biased against women \citep{Amazon}. For example, the algorithm penalized resumes with the word `women’s’ in it (e.g., `women's cross country team captain' or `women's union staff member') as well as graduates from two all-women’s colleges \citep{Amazon}. Amazon's algorithm seems unfair to female applicants. Namely, the algorithm wrongfully discriminated against female applicants by using a success metric (i.e., the similarity between the applicant’s resume and previously submitted resumes) that systematically overlooks well-qualified female applicants in virtue of their gender identity. However, as previously discussed in Section \ref{subsec:data-science-ethics}, it is unclear how to evaluate fairness in models like Amazon's now-scrapped hiring algorithm and, with that, how exactly to mitigate any unfairness in such models. 

Moreover, it seems conceivable that even if Amazon improved the success metric, its hiring algorithm would still be biased against female applicants in virtue of there being a gender bias in the training data (i.e., model's input). The saying ``garbage in, garbage out” encapsulates the commonly seen connection between bias, fairness, justice, and data modeling. That is, the saying ``garbage in, garbage out” notes that if the data used to build the data model was biased against group X, then the model’s predictions would be biased against group X and could lead to unfair (and/or unjust) outcomes for group X if the model's predictions influence decision-making.

Beyond data modeling, issues of bias, fairness, and justice also emerge during ``interactions with the world". For instance, suppose a data scientist wants to model the average number of hours in the hospital after giving birth but only surveys white females. We would consider the dataset biased towards white women and consequently be cautious about generalizing the data scientist’s findings to people in the population who are not white women.

\paragraph{Less conventional example(s):} However, issues of bias, fairness, and justice are crucial to consider at every stage of the data science lifecycle. For example, we might completely drop observations with missing values when processing the data. Yet, dropping those values can create biases in our data and subsequent analyses if they are not missing at random.\footnote{This example is partially inspired by the discussion in \citet{non-random-imputation}.} Bias, fairness, and justice can also come into play when insights from a data model influence our future interactions with our world. For instance, it seems unfair (and/or unjust) to only give a nannying job ad to women because an algorithm found that women were substantially more likely than men to click on nannying job ads.

\subsubsection{\href{https://scolando.github.io/data-science-ethics/readings/Explainability.html}{Explainability, Interpretability, Transparency}}

A common complaint about data models, particularly more advanced ones, is that they are black boxes. In this subsection, we focus mainly on the connections between the lifecycle and explainability. However, the connections are similar to those for transparency and interpretability, given all three concepts share the common goal of making data models and their predictions more understandable to stakeholders.

\paragraph{Paradigmatic example(s):} In the model deployment stage, where predictions about individuals or groups are made, a reasonable ethical expectation is that the model is understandable to human stakeholders. One reason for this expectation is that if the model is explainable, then we can verify that the model is fair. For example, if a convicted person’s bail is set (using an algorithmic recommendation) higher than they think it should be, it seems fair for the individual to demand and expect an explanation for the algorithm's recommendation in order to ensure that sensitive social attributes like race, gender, or socioeconomic status did not impact the algorithm's prediction. Similarly, stakeholders might also demand that the algorithm's parameters be transparent in order to see which variables influence the model's predictions and, in particular, see whether the algorithm uses sensitive social attributes to make its predictions. The ethical expectation that data models are understandable to human stakeholders is often referred to as the ``right to an explanation." This expectation is echoed in legal documents such as the European Union's General Data Protection Regulation (GDPR), which states that an individual has the right to ``obtain an explanation" for any automated decision made about them \citep{Goodman2017}.

\paragraph{Less conventional example(s):} While not mentioned as often, the ``right to an explanation” is also important to consider when insights from a data model influence future interactions with the world. For instance, suppose that law enforcement starts heavily policing Neighborhood A relative to Neighborhood B because a data model found that the people in Neighborhood A are more likely to be convicted of a crime than the people in Neighborhood B. It seems that the people in Neighborhood A have a right to understand why they are being policed more than the people in Neighborhood B. Like in the paradigmatic example, one reason for why the people in Neighborhood A seem to have a ``right to an explanation" in this case is that explainability is often necessary to ensure that the model does not use sensitive social attributes to make its predictions. However, some philosophers note that human decision-makers are also black boxes with respect to how they arrive at their decisions. For instance, even if a court judge gives justification for their sentencing, it is not clear that the reason they give is the only reason for their decision or even is a reason for their decision at all (e.g., people can be subconsciously influenced by implicit biases) \citep{Gunther2022}. As such, one question within philosophy is whether we have a ``right to an explanation" when it comes to data models, and if so, why do we have such a ``right to an explanation" when it seems to hold algorithms to a higher standard than human decision-makers? 

\subsubsection{\href{https://scolando.github.io/data-science-ethics/readings/Causation.html}{Causation}}

\paragraph{Paradigmatic example(s):} As early as introductory statistics courses, ``correlation does not imply causation" is emphasized. The topic of causation is prominent in many data science ethics courses because mistakenly claiming causation can be morally pernicious when attempting to interpret knowledge from a data model. Most current data models can only identify correlations between the predictor variables and the response variable rather than causal relationships (n.b., there are also centuries worth of philosophical debates about how to define causation).\footnote{Also, see examples of how current data models are not yet able to accurately identify causation in \citet{towardsdatascienceCausalDiscovery}.}

Incorrectly interpreting causation between variables can have substantial moral repercussions during model deployment. For instance, suppose we have a logistic model that predicts whether a person will drop out of high school. Our model finds a positive association between having a first language other than English and the expected probability of dropping out of high school. There are several confounding variables, like socioeconomic status and availability of academic opportunities, which explain the identified positive association. However, imagine that a person sees our model and, from it, concludes that having a first language other than English {\em causes} a higher probability of dropping out of high school. As a result of mistakenly drawing causal claims from the model, they might advocate for English-only policies or develop prejudiced beliefs against people whose first language is not English.

\paragraph{Less conventional example(s):} Moral implications surrounding causation can also arise when building causal inference models. For example, imagine there is a group of researchers who want to understand how a patient's race influences their wait time in an emergency room. Answering the wait time question would involve evaluating whether the following counterfactual is true: if a patient was a member of racial group X instead of racial group Y, then their ER wait time would be different. One way the researchers might try to evaluate this counterfactual is by collecting data that contains background information (e.g., race) and the wait time at the hospital for each patient. The data would be used to estimate how changing only the `race' variable for a patient would change their wait time. Yet, \citet{causal-inference-worries} note that changing only the `race' variable endorses an essentialist view of race that fails to acknowledge how race is socially constructed. Essentialist views of variables like race are harmful because they ignore the complex social, historical, and political factors that shape individuals' experiences. As a result, essentialist views can often lead to overgeneralizations about social groups and even wrongful discrimination against them \citep{essentialism-harm}. Worries about how to thoughtfully conduct causal inference on social categories point to the importance of reflecting on philosophical questions, like what constitutes a specific social category (e.g., race, gender, sexuality, etc.) during model building.

\subsubsection{\href{https://scolando.github.io/data-science-ethics/readings/Democracy.html}{Democracy}, 
\href{https://scolando.github.io/data-science-ethics/readings/Workplace.html}{Workplace}, \href{https://scolando.github.io/data-science-ethics/readings/Predictive-Policing.html}{Predictive Policing}}
\label{case-studies}

\paragraph{Paradigmatic example(s):} Democracy, workplace, and predictive policing are all settings where data models are used and have exceptionally high moral stakes. As such, case studies related to democracy, workplace, and predictive policing are very common in the collated data science ethics syllabi and reading lists. Usually, when democracy, workplace, and predictive policing case studies are referenced, the focus is on model deployment. For instance, COMPAS is often brought up because it is deployed in a setting with high moral stakes, i.e., in US courtrooms to aid judges in making decisions about bond amounts and sentencing lengths for defendant \citep{COMPAS}.

Yet, while there are moral implications for deploying data models in democracy, workplace, and predictive policing settings, it does not seem that creating such data models is inherently morally problematic. For example, suppose that a civil rights group creates an algorithm to predict a defendant's likelihood of being convicted (like COMPAS does) but only uses the model to show that the criminal justice system is racially biased against Black defendants. Intuitively, building such a data model is not morally problematic. Rather, data models that predict a defendant's recidivism risk are morally problematic when they are deployed in such a way that the model's predictions impact people's beliefs about a defendant's recidivism risk and court outcomes. 

The ethical theory of {\em consequentialism} can explain why the model's deployment is relevant to its moral evaluation. According to {\em consequentialism}, only the consequences of an action ought to influence our moral assessment of it \citep{StanConsequentialism}.  In the civil rights group and COMPAS examples, the consequences are different. COMPAS is being used to set bond amounts, sentence length, and parole, whereas the civil rights group algorithm is not being used in such a capacity. Another relevant difference between the civil rights group algorithm and COMPAS is that the civil rights group algorithm works {\em against} existing injustice rather than compounding it by aiming to elucidate the existing racial bias within the criminal justice system. Philosophy helps locate what exactly is morally problematic in a specific case (e.g., is it the data model's predictions in themselves or how the model is deployed?), thereby helping us make our data science practices more ethical.

\paragraph{Less conventional example(s):} Interpreting knowledge from model predictions can also lead to morally problematic interactions with the world within democracy, workplace, and predictive policing settings. For example, in her book, {\em Weapons of Math Destruction}, Cathy O'Neil considers PredPol, an algorithm that uses historical crime data to predict where crimes are most likely to occur. When police use PredPol, they can target neighborhoods based on where ``nuisance" crimes (e.g., vagrancy, aggressive panhandling, selling and consuming small quantities of drugs) occur, which are unlikely to be recorded when there is not a police officer present \citep{oneil2016}. However, \citet{oneil2016} also notes that ``nuisance" crimes are also much more common in impoverished neighborhoods. When the	``nuisance" crime data is put into the predictive model, more police patrol impoverished neighborhoods, and arrests in those neighborhoods are more likely to occur. This creates a pernicious feedback loop because the policing of impoverished neighborhoods leads to arrests in the neighborhood, which ultimately justifies more policing of those neighborhoods \citep{oneil2016}. As a result, more people are arrested for ``nuisance" crimes, the majority of which come from impoverished neighborhoods and are Black or Hispanic due to racial segregation in cities \citep{oneil2016}. Thus, using predictive policing models to inform future interactions with the world (i.e., where to send police) can create pernicious feedback loops that exacerbate existing racial injustices in the criminal justice system.

\section{Discussion}
\label{sec:discussion}

Interdisciplinary teaching is hard. Not only does it require double the amount of knowledge, but it also requires instructors to think differently. Arguments made in data science are very different from those made in philosophy. Even the words (e.g., {\em fairness}) often take on different meanings across the two disciplines. In order to teach data science ethics well, we recommend that you embrace what each of the two disciplines has to offer and try to deliberately bring each into your classroom. 

For example, consider the data science ethics issues surrounding COMPAS. A data scientist might approach the ethical problem of whether or not the algorithm is fair by noting that it is impossible to simultaneously have {\em equal false positive rates}, {\em false negative rates}, and {\em predictive parity} across Black and white defendants \citep{chouldechova2017,kleinberg2016, Corbett-Davies}. But what does {\em fairness} even mean? \citet{Castro2022} argues that an algorithm is fair if and only if it does not wrongfully discriminate against members of protected classes (e.g., members of a certain racial group or gender). Per \citet{Castro2022}, statistical criteria, which equalize summary statistics across groups, do not capture the broad complaints we have about algorithmic fairness. Specifically, statistical criteria fail to evaluate whether the algorithm is sensitive to the needs of those who are worse off, places burdens arbitrarily on certain individuals but not others, and perpetuates unjust social structures \citep{Castro2022}. In contrast, \citet{hellman2019} contends that statistical criteria can be helpful when evaluating algorithmic unfairness. Specifically, having an unequal ratio of false positives to false negatives between groups, while not constitutive of unfairness itself, is important to examine when evaluating whether an algorithm is unfair \citep{hellman2019}. Meanwhile, \citet{vredenburgh2023} argues that {\em fairness} is not even a relevant concept for judging algorithms when they are deployed within systems with pre-existing injustice because what we ought to prioritize in such scenarios is {\em justice}. Alternatively, \citet{johnson2021} looks at issues of {\em algorithmic bias} and their connections to {\em cognitive bias}, specifically in the context of the COMPAS algorithm. Johnson highlights that the similarities between algorithmic and cognitive biases underscore that algorithmic biases need not necessarily be from explicitly biased code. Rather, algorithmic biases can arise from ``seemingly innocuous patterns of information processing," which often can make the bias harder to mitigate \citep{johnson2021}. At the intersection of ethics and data science, \citet{engel2024} argue that normative questions, for example, how we should weigh the false positives against the false negatives, are baked into COMPAS' recidivism code.  In a data science class that discusses the COMPAS algorithm, students should be exposed to literature from both data scientists and philosophers. A conversation about not only the two sets of arguments but also about the ways in which the arguments {differ} from one another will deepen the student's understanding of the ethical considerations at hand. Indeed, a potential assignment could be for students to choose one data science and one philosophical argument to present to the class. A discussion of the myriad of approaches to evaluating algorithmic fairness would be engaging and enlightening for the students.

In addition to engaging with cross-disciplinary literature, we recommend other ways to infuse philosophical ideas into a data science ethics class. If possible, find allies from the other discipline with whom to co-teach. If not possible, strike up conversations (have lunch!) with allies who are equally interested in data science ethics topics and have a different academic background from your own. Organize your syllabus topics with a lens toward explicitly connecting ethics and data science. To find such connections, we suggest using the examples outlined in Section \ref{sec:connections}, referring to the reading lists on our website (linked in each subsection's header in Section \ref{sec:connections}), or finding others who have taught in the area and already identified such connections. Use case studies as a starting point\footnote{The data science ethics case studies from the Markkula Center for Applied Ethics are excellent \citep{santaclara}.}, then zoom out to consider philosophical concepts that overlap with the ideas being covered to grapple more directly with the difficult philosophical questions that arise in data science (\usebibentry{SEP}{title} is a fantastic reference for learning more about philosophical concepts \citep{SEP}). Suggest to students that they focus on how the case study relates to a particular philosophical concept instead of focusing on whether the algorithm is ``good" or ``bad," all things considered. For instance, instead of asking whether the COMPAS algorithm is bad to use, ask what exactly COMPAS teaches us about algorithmic bias and fairness. For one, what questions does it prompt about formal definitions of algorithmic fairness, like {\em equal false positive rates} or {\em predictive parity}?  Step outside your comfort zone, work toward engaging with a new way of thinking, and encourage your students to do the same.

\section{Conclusion}
\label{sec:concl}

Our motivation for this project was to investigate the connections between data science frameworks and ethical frameworks in existing data science ethics courses. Our investigation demonstrated that data science is rife with potentially morally charged decision points, as depicted in Figures \ref{fig:DS-stack-values} and \ref{fig:lifecycle_2}. The extensive connections between common ethics topics and the data science lifecycle explained in Section \ref{sec:connections} highlight that embracing the large scholarship in philosophical ethics, which has been developed over thousands of years, is integral to understanding what potentially morally charged decisions exist in data science and how we ought to navigate them. 

Our \href{https://scolando.github.io/data-science-ethics/}{data science ethics website} contains much of the content provided in this paper. However, it expands the work to include course descriptions for the sixteen data science ethics classes we examined and reading lists of papers for each ethics topic described in Section \ref{sec:connections}.

We hope that the connections we have made between data science and philosophical ethics, as well as our created website, will generate conversation and encouragement to make deeper connections in the classroom between ethics and data science.

\section*{Acknowledgments}

We would like to thank Brad McHose for helpful input and feedback on an early draft of this manuscript. We also thank the Pomona College SURP program and Kenneth Cooke Summer Research Fellowship for supporting SC in summer research.  

\section*{Data Availability Statement}

The work (the dashboard) that supports the findings of this study are openly available at \url{https://scolando.github.io/data-science-ethics/}.  The syllabus topics are available as a .csv file at \url{https://osf.io/8q2me/}.

\begingroup
\raggedright

\bibliographystyle{apalike}

\bibliography{ds_ethics.bib}
\endgroup

\appendix
\section{Appendix}
\label{app-rep}

In the appendix, we provide worked-out examples and describe some of the limitations of the lifecycle we used in this paper to underscore the value of judgment in defining relationships between data, the world, and knowledge. We hope that our examples lead to interesting and fruitful classroom discussions.

Even though, as highlighted in Section \ref{relational}, it makes the most sense to think of data and data models using a relational view, it is worth noting (and pointing out to students) that there exist alternative frameworks in which to consider data and data models. Here, we examine the representational view of data, data models, and their connections to what we interpret as knowledge about our world \citep{Leonelli}.

In Sections \ref{representational} and \ref{relational}, we explain the representational and relational view of data and data models, respectively. In Section \ref{rep-rel-in-practice}, we provide two examples to demonstrate why the relational view of data and data models is preferable to the representational view.\footnote{We recommend reading \citet{Leonelli} as well as sections of \citet{Beaulieu-Leonelli} for a more in-depth justification for why the relational view of data and data models is preferable to the representational view.} 

\subsection{The Representational View}
\label{representational}

\begin{figure}[H]
\centering
\includegraphics[scale = 0.4]{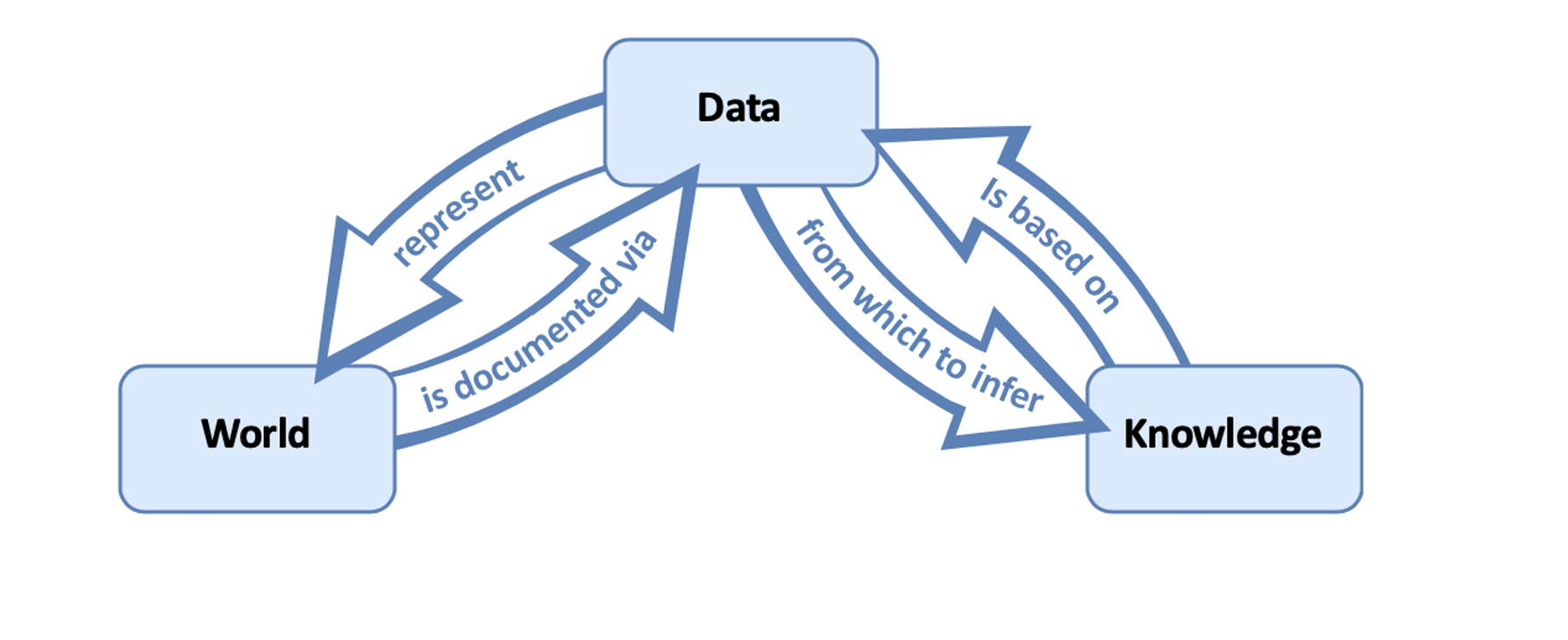}
\caption{Data science lifecycle under the representational view of data and models as formulated by \citet{Leonelli}.}
\label{fig:rep_lifycycle}
\end{figure}

Under the representational view of data and data models (see Figure \ref{fig:rep_lifycycle}), the informational content of data is fixed and independent of the researchers’ background assumptions and context. Thus, data models are only important insofar as they extract the underlying truth from the data. So, under the representational view, data models are either correct or incorrect, depending on their ability to elucidate the truth stored in the data. In other words, data are simply numbers that hold information, and data models (and other methods of processing data) are only relevant because they clarify the connections between data and what we interpret to be knowledge \citep{Leonelli}.\footnote{Note that what we interpret as knowledge is {\em not} the same as actual knowledge. E.g., the statement ``the Earth is flat" was historically interpreted as knowledge, even though it is a false statement and thus not actual knowledge.}

\subsection{The Relational vs. Representational View In-Practice}
\label{rep-rel-in-practice}

\paragraph{Predicting the Likelihood of Buying Concert Tickets}

Suppose we are interested in predicting a person's likelihood of buying concert tickets from a particular website. To do so, we collect data about the number of times they clicked on an advertisement for concert tickets from that particular website, the timestamps of these ad clicks, the person's demographic information, et cetera.

However, it is unclear what exactly the data we gather should be taken to represent. We concede that we cannot directly measure a person's interest in buying concert tickets, but we believe that someone's interest is correlated to how often they interact with online ads for the tickets. So, we decide to use a person's number of ad clicks as a proxy for their interest in buying concert tickets. However, this proxy is imperfect. For instance, a person might click on a ticket ad in order to determine how much to resell their concert tickets for.

Furthermore, using particular data as evidence can sometimes influence future interactions with the world. Imagine that we find that when a website displays, ``less than 1\% of tickets remaining", the person is much more likely to buy concert tickets. As a result, other ticket sites adopt this strategy in an attempt to sell more tickets. However, suppose there is only a strong correlation between displaying the message ``less than 1\% of tickets remaining" and a person's likelihood of buying tickets when no other site displayed a similar message. Namely, when other sites also display the message ``less than 1\% of tickets remaining," displaying the message on our site will no longer increase the person's likelihood of buying tickets from our site. What this example demonstrates is that by treating the display message data as evidence for there being a correlation between displaying the message and the likelihood of buying a ticket, we have changed how future users will interact with our site.  

Unlike the representational view, the relational view acknowledges that data's informational content is influenced by researchers' background assumptions and social contexts. Furthermore, the relational view endorses that data can be dynamic, and what we take as evidence can influence future interactions with the world. Thus, the concert ticket example described above gives us reason to endorse the relational view of data and data models over the representational view. 

\paragraph{Predicting the Likelihood of Giving a Red Card}

In \citet{soccer-case}, 29 data analysis teams were asked to use the same data set to determine ``whether soccer referees are more likely to give red cards to dark-skin-toned players than light-skin-toned players." Despite operating from the same data set, the final conclusions were split: 20 teams found that there was a statistically significant positive relationship, and 9 teams did not find a significant association between skin tone and the likelihood of the referee giving a red card. 

The difference in the chosen data model type and the relative importance of the potential predictor variables contributed to the division in the teams’ findings:

\begin{itemize}
  \item 15 teams used logistic models, 6 teams used Poisson models, 6 teams used linear models, and 2 teams used other types of models.
  \item 21 of 29 teams used unique combinations of predictor variables.
\end{itemize}

Through \citet{soccer-case}, we can also see how ambiguity about the data model and the relative importance of certain predictor variables impacts what data is taken as evidence. No two teams had the same set of evidence for their claim about the relationship between skin tone and the likelihood of the referee giving a red card. As emphasized by \citet{soccer-case}, each team’s evidence set was defensible based on the original data set provided. Yet, these evidence sets were also subjective in the sense that they relied upon the analysts’ background assumptions, value judgments, and social contexts. These subjective factors ultimately influenced what the analysts interpreted as the true relationship between the player's skin tone and the likelihood of the referee giving the player a red card. Hence, the \citet{soccer-case} case study emphasizes that data and data models should be viewed relationally rather than representationally.

\subsection{}
\label{flaws-lifecycle}

For instructors who seek to walk their students through the advantages and disadvantages of specific data science lifecycles, we note some flaws with our selected lifecycle (see Figure \ref{fig:lifecycle-practices}). First, the arrows between the data science stages imply that the data science is linear (i.e., only when all prior steps are complete is the next step pursued). However, in practice, data science is iterative; for example, data scientists may build a data model, realize their model underperforms on a certain group, and then collect more data to improve the model's performance. Another shortcoming of our final lifecycle in Figure \ref{fig:lifecycle-practices} is that it does not highlight how many different people (or groups) are often involved in different stages of the lifecycle. For instance, a company might collect the data and then bring in an external group of data scientists to create a meaningful model from that data. The fact that different people (or groups) act at different stages of data science is essential both for understanding how data science practices are actually carried out as well as why understanding certain ethics topics (such as moral responsibility) is integral to data science. As such, we move forward with our final data science lifecycle in the spirit of George E. P. Box, ``all models are wrong, but some are useful." \citep{box1979}

\end{document}